\documentclass[pra,showpacs,superscriptaddress,nobalancelastpage]{revtex4}
\usepackage{graphicx}
\usepackage{amsmath}
\usepackage{amsfonts}
\usepackage{amssymb}
\newtheorem{proposition}{Proposition}
\newenvironment{proof}[1][Proof]{\textbf{#1.} }{\ \rule{0.5em}{0.5em}}

\begin{document}

\title{Conditioning two-party quantum teleportation within a three-party quantum channel}
\author{Stefano Pirandola}
\author{Stefano Mancini}
\author{David Vitali}
\affiliation{Dipartimento di Fisica, Universit\`a di Camerino, I-62032 Camerino, Italy}

\begin{abstract}
We consider an arbitrary continuous variable three-party Gaussian
quantum state which is used to perform quantum teleportation of a
pure Gaussian state between two of the parties (Alice and Bob). In
turn, the third party (Charlie) can condition the process by means
of local operations and classical communication. We find the best
measurement that Charlie can implement on his own mode that
preserves the Gaussian character of the three-mode state and
optimizes the teleportation fidelity between Alice and Bob.
\end{abstract}


\maketitle

\section{Introduction}

Quantum information processing with continuous variables (CV)
provides an interesting alternative to the traditional qubit-based
approach. CV seem to be particularly suitable for quantum
communications, as for example quantum teleportation \cite{telepo}
and quantum key distribution \cite{QKD}. Multipartite CV entangled
states for quantum communication networks are rather easy to
produce. In particular, tripartite entangled Gaussian states
realized either using squeezers and beam splitters \cite{Braun},
interlinked bilinear interactions \cite{parisopto} or through
radiation pressure \cite{PRL,PRA,network,JOBSO} have been
considered for the realization of a quantum teleportation network
and for telecloning. Tripartite CV entanglement between optical
modes has been generated using squeezing and beam splitters and
experimentally characterized in \cite{cinjap}, and it has been
recently exploited for the realization of quantum secret sharing
in \cite{Lance} and for quantum telecloning in \cite{FuruNat}.
Here we consider a generic CV tripartite Gaussian state which is
employed for the specific task of teleporting a pure Gaussian
state between two of the three parties (Alice and Bob). We
determine the best way the third party (Charlie) can cooperate to
improve this teleportation task. To be more specific, we find the
optimal Gaussian measurement at Charlie's site which maximizes the
teleportation fidelity. This is different from the optimization
over all possible local Gaussian operation of CV teleportation as
considered in \cite{fiurasek} and also from the problem of
entanglement distillation \cite{distill}, where one always starts
from \emph{bipartite} entangled states and tries to increase their
entanglement. In Sec. II we present the scenario and describe the
teleportation protocols in the case when it is assisted or is not
assisted by measurements at Charlie's site. In Sec. III we discuss
the case when Charlie performs a dichotomic measurement with a
Gaussian and a non-Gaussian outcome, while in Sec. IV we consider
the case of a local Gaussian measurement performed at Charlie's
site. In Sec. V various applications of the theorems derived in
Sec. III and IV are discussed in detail, while Sec. VI is for
concluding remarks.

\section{Assisted and non-assisted teleportation protocols}

The scheme we are going to study is described in Fig.~1: Alice,
Bob and Charlie each possess a continuous variable mode,
characterized by an annihilation operator $\hat{a},\hat{b}$ and
$\hat{c}$ respectively, and share a quantum channel given by an
arbitrary three-mode Gaussian state $\rho$, characterized by a
displacement $\vec {d} \in {\mathbb R}^6$ and a correlation matrix (CM)
\begin{equation}
V\equiv\left(
\begin{array}
[c]{ccc}
A & F & E\\
F^{T} & B & D\\
E^{T} & D^{T} & C
\end{array}
\right),  \label{CMtot}
\end{equation}
where the blocks $A,B,...,F$ are $2\times2$ real matrices. Alice
has to teleport to Bob an unknown pure Gaussian state $\rho_{in}$ with CM $V_{in}$ and amplitude
$\mu$. The most straightforward strategy is to ignore Charlie
(non-assisted protocol) and use the reduced bipartite Gaussian
state $\rho^{tr}\equiv {\rm Tr}_{c}(\rho)$ to implement a
standard continuous variable teleportation protocol \cite{telepo}.
In such a case, Alice mixes her part of the reduced state
$\rho^{tr}$\ with the input state $\rho_{in}$ through a balanced
beam-splitter and makes a homodyne detection of the output modes,
i.e., she measures the quadratures
$\hat{X}_{-}\equiv2^{-1/2}(\hat{X}_{a}-\hat{X}_{in})$ and
$\hat{P}_{+} \equiv2^{-1/2}(\hat{P}_{a}+\hat{P}_{in})$. After the
measurement, Alice classically communicates the result $\gamma
\equiv-X_{-}+iP_{+}$ to Bob, who performs a conditional
displacement
$\hat{D}_{b}(\gamma^{\prime})\equiv\exp(\gamma^{\prime}\hat{b}^{\dagger}
-\gamma^{\prime\ast}\hat{b})$ on his own mode $b$, where
$\gamma^{\prime }=\gamma+\delta$ has the double effect to
compensate the displacement due to Alice's measurement (by $\gamma$) and the
displacement of the reduced state (by $\delta$ which is connected with
$\vec{d}$, see \cite{network}). This means that Bob can always
implement (through a suitable displacement $\gamma^{\prime}$) a
displacement-independent teleportation protocol, whose fidelity only
depends on the CMs of the reduced state and the input state.
When Charlie is traced out, $\rho^{tr}$ has the CM
\begin{equation}
V^{tr}=\left(
\begin{array}
[c]{cc}
A & F\\
F^{T} & B
\end{array}
\right)  \label{CMtr}
\end{equation}
and the teleportation fidelity is given by $F^{tr}=(\det\Gamma^{tr})^{-1/2}$ \cite{fiurasek}, where
\begin{equation}
\Gamma^{tr}\equiv2V_{in}+RAR+B-RF-F^{T}R \label{gammatr}
\end{equation}
and $ R={\rm diag}\left(1,-1\right)$.

\begin{figure}[ptbh]
\vspace{-0.6cm}
\begin{center}
\includegraphics[width=0.65\textwidth]{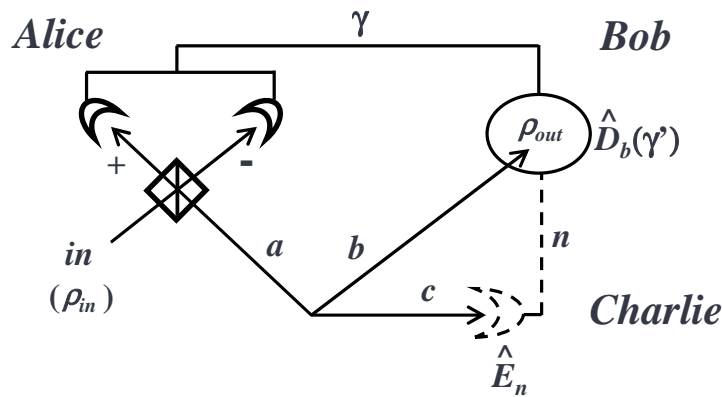}
\end{center}
\vspace{-1.6cm}\caption{An arbitrary 3-mode Gaussian state is
shared by Alice, Bob and Charlie. Alice is supplied with an
unknown pure Gaussian state $\rho_{in}$ which she wants to
teleport to Bob. In a first strategy, Charlie is traced out and
Alice and Bob implement a standard continuous variable
teleportation protocol. In an alternative strategy (dashed
detector), Bob is helped by Charlie who measures his mode and
classically communicates the result $n$ to Bob, who uses also this
information for his local operation. Here we consider, for
Charlie's measurement, first a local dichotomic measurement and
then a local Gaussian measurement.} \label{setup}
\end{figure}

An alternative strategy for Alice and Bob is to ask for the help
of Charlie (assisted protocol), who can perform a suitable
measurement on his own mode $c$ and classically communicate the
result to Bob (see Fig.~\ref{setup}). In this modified protocol,
Bob performs his displacement only after receiving the information
about the measurement outcomes from {\em both} Alice and Charlie.
For every Charlie's outcome $n$ (with probability $P_{n}$), Bob
can choose a conditional displacement $\gamma_{n}^{\prime}$ aiming
at optimizing the conditional fidelity $F^{(n)}$ and therefore the
effective fidelity $F=\sum_{n}P_{n}F^{(n)}$ of the protocol
\cite{conditional}. In particular, if the bipartite reduced state
conditioned to the outcome $n$, $\rho^{(n)}$, is a Gaussian state,
then Bob's displacement is given by
$\gamma_{n}^{\prime}=\gamma+\delta_{n}$\ where $\delta_{n}$
exactly cancels the displacement of $\rho^{(n)}$, and therefore
the conditional fidelity $F^{(n)} $\ depends, as before, only on
the CMs. In the following we consider two general kinds of
measurement at Charlie's site: a local dichotomic measurement,
with a Gaussian outcome and a non-Gaussian one, and a local
Gaussian measurement, defined as a local measurement preserving
the Gaussian character of the shared state for every outcome. Our
aim is to compare the assisted fidelity $F$ and the non-assisted
fidelity $F^{tr}$ for both kinds of measurement. We anticipate
that for the dichotomic measurement one does not have an
improvement ($F\leq F^{tr}$), but the results achieved for the
conditional fidelities $F^{(n)}$\ are interesting and they can be
directly extended to the case of the Gaussian measurement where
one can optimize $F$ and then surely state that $F\geq F^{tr}$.

\section{Dichotomic measurement}

We first consider the case of a dichotomic measurement with
measurement operators
$\hat{E}_{0},\hat{E}_{1}\equiv(\hat{I}-\hat{E}_{0}^{2})^{1/2}$
where $\hat{E}_{0}$ is an arbitrary Gaussian state with
displacement $\vec{d}_{0}$ and CM $V_{0}$. This implies that for
the outcome $n=0$ the conditional bipartite state
$\rho^{(0)}\equiv P_{0}^{-1}{\rm
Tr}_{c}(\hat{E}_{0}\rho\hat{E}_{0}^{\dagger})$ is still Gaussian,
while for the other outcome it is not Gaussian. One can prove (see
Appendix A) that the CM of the reduced state $\rho^{(0)}$ is
\begin{equation}
V^{(0)}=V^{tr}-\left(
\begin{array}
[c]{cc}
EME^{T} & EMD^{T}\\
DME^{T} & DMD^{T}
\end{array}
\right)  \label{CM0}
\end{equation}
where $M$ is the following 2$\times$2 ``measurement matrix''
\begin{equation}
M\equiv\frac{1}{g}\Omega\left[2(\det V_{0}+1/4)V_{0}+4(\det
V_{0})C\right]\Omega^{T} \label{M}
\end{equation}
with \begin{equation} g\equiv4\det V_{0}\det C+2(\det
V_{0}+1/4){\rm Tr}(V_{0}\Omega C\Omega^{T})+(\det V_{0}+1/4)^{2}>0
\label{fattoreG} \end{equation} and
\begin{equation}
\Omega\equiv\left(
\begin{array}
[c]{cc}
0 & -1\\
1 & 0
\end{array}
\right)  .\label{Omlabel} \end{equation} If we use Eq.~(\ref{CM0})
in the right hand side of Eq.~(\ref{gammatr}) instead of
Eq.~(\ref{CMtr}), we obtain
\begin{equation}
\Gamma^{(0)}=\Gamma^{tr}-\Sigma^{T} M\Sigma \label{gamma0}
\end{equation}
where
\begin{equation}
\Sigma\equiv E^{T}R-D^{T}  . \label{sigma}
\end{equation}
Now, the conditional teleportation fidelity is given by
$F^{(0)}=(\det \Gamma^{(0)})^{-1/2}$ and satisfies the following

\begin{proposition}
The conditional fidelities $F^{(n)}$ corresponding to the outcomes
$n=0,1$ and the assisted fidelity $F$\ satisfy the inequality
\begin{equation}
F^{(1)}\leq F\leq F^{tr}\leq F^{(0)}\label{result1}
\end{equation}
\end{proposition}

\begin{proof}
The proof of $F^{tr}\leq F^{(0)}$ is based on Eq.~(\ref{gamma0}).
Matrix $M$ is real, symmetric and strictly positive and, since
$\Sigma$ is real, the matrix $\Sigma^{T}M\Sigma$ is real,
symmetric and positive. Likewise $\Gamma^{tr}$ and $\Gamma^{(0)}$\
are real, symmetric and strictly positive, and from linear algebra
it follows that $\det(\Gamma^{(0)})=\det(\Gamma
^{tr}-\Sigma^{T}M\Sigma)\leq\det(\Gamma^{tr})$. The proof of
$F^{(1)}\leq F^{tr}$ is a consequence of the previous result. The
characteristic functions $\Phi^{tr}$, $\Phi^{(0)}$ and
$\Phi^{(1)}$ of states $\rho^{tr}$, $\rho^{(0)}$ and $\rho^{(1)}$
are related by $\Phi^{(1)}=P_{1}^{-1}[\Phi^{tr}-P_{0} \Phi^{(0)}]$
which also shows that $\rho^{(1)}$\ is not Gaussian. On the other
hand, the fidelity $F^{(1)}$ can be expressed in the form
\cite{network} \ $F^{(1)}=\pi^{-1}\int d^{2}\lambda\left|
\Phi_{in}(\lambda)\right|
^{2}[\Phi^{(1)}(\lambda^{\ast},\lambda)]^{\ast}\exp(-\lambda\delta^{(1)\ast
}+\lambda^{\ast}\delta^{(1)})$ where $\Phi_{in}$ is the
characteristic function of the input state and $\delta^{(1)}$ is
an additional shift optimizing $F^{(1)}$. Now, it is easy to prove
that, for every $\delta^{(1)}$, one has $F^{(1)}\leq
P_{1}^{-1}[F^{tr}-P_{0}F^{(0)}]\leq P_{1}^{-1}
[F^{tr}-P_{0}F^{tr}]=F^{tr}$. Finally, the effective fidelity is
given by $F=P_{0}F^{(0)}+P_{1}F^{(1)}$ and, since $F^{(0)}\geq
F^{(1)}$, one has $F\geq F^{(1)}$. On the other hand, from
inequality $F^{(1)}\leq P_{1}^{-1} [F^{tr}-P_{0}F^{(0)}]$\ one can
derive $F\leq P_{0}F^{(0)}+[F^{tr} -P_{0}F^{(0)}]=F^{tr}$.
\end{proof}

Eq.~(\ref{result1}) shows that the dichotomic measurement leads,
on average, to an assisted fidelity $F$ which does not outperform
$F^{tr}$, proving that the present dichotomic scheme does not seem
to bring advantages in a real teleportation network. However the
situation is very interesting from the point of view of the
conditional teleportation fidelities $F^{(n)}$. In fact
Eq.~(\ref{result1}) shows that teleportation fidelity always
increases if Charlie performs a measurement and the corresponding
conditional state is still Gaussian ($F^{(0)}\geq F^{tr}$), while
it always decreases with respect to the trace case for the outcome
corresponding to the non-Gaussian conditional state ($F^{(1)}\leq
F^{tr}$), even if in this latter case the conditional bipartite
state is more pure than that without measurement \cite{Chuang}.
This result suggests which is the right kind of measurement to be
considered at Charlie's site (local Gaussian measurement) and it
will be the starting point of the next section~IV.

Moreover, the dichotomic scheme can be used in a probabilistic
way, i.e., selecting only the Gaussian outcome. In this case Bob
asks Alice to perform the Bell measurement and the classical
communication only if Charlie's measurement has given the Gaussian
outcome. In such a case the assisted fidelity $F$ is just the
conditional one $F^{(0)}$, but the protocol has a success
probability equal to $P_0$.

We have also explicitly
verified (see section~V.B) that the Gaussian outcome $n=0$ can
give $F^{(0)}>1/2$ for the teleportation of coherent states when
$\rho^{tr}$ is not entangled (and therefore $F^{tr}\leq1/2$). In
other words, Charlie can conditionally generate remote bipartite
entanglement between Alice and Bob if the dichotomic measurement
selects the Gaussian outcome. All these considerations make clear
why it is profitable to optimize the ``Gaussian'' conditional
fidelity $F^{(0)}$ upon the measurement parameters and exactly
such optimization work concerns the remainder of this section.

Thus we restrict to the Gaussian outcome ($n=0$), and look for the
optimal Gaussian state $\hat{E}_{0}$ (i.e. the optimal CM $V_{0}$)
which maximizes the fidelity $F^{(0)}$. As a first result we can
prove the following

\begin{proposition}
For every Gaussian state $\hat{E}_{0}$, there exists a pure Gaussian state
$\hat{E}_{0,p}$ such that $F^{(0,p)}\geq F^{(0)}$.
\end{proposition}
\begin{proof}
For every Gaussian state $\hat{E}_{0}$, there exists a Gaussian
unitary transformation $\hat{U}$ such that
$\hat{E}_{0}=\hat{U}\rho(n_{T})\hat{U}^{\dagger}$ where
$\rho(n_{T})$ is a thermal state with $n_{T}\geq0$ mean number of
photons \cite{duan}. Thus, we can rewrite the reduced state
$\rho^{(0)}\equiv P_{0} ^{-1}{\rm
Tr}_{c}(\hat{E}_{0}\rho\hat{E}_{0}^{\dagger})=P_{0}^{-1}{\rm
Tr}_{c}[\rho (n_{T})\hat{U}^{\dagger}\rho\hat{U}\rho(n_{T})]$ so
that the fidelity $F^{(0)}$ achieved from the tripartite Gaussian
state $\rho$ and the measurement operator $\hat{E}_{0}$ is the
same which is achieved from the tripartite Gaussian state
$\hat{U}^{\dagger}\rho\hat{U}$ and the measurement operator
$\rho(n_{T})$ i.e. $F^{(0)}\equiv
F(\rho,\hat{E}_{0})=F[\hat{U}^{\dagger}\rho\hat{U},\rho(n_{T})]$.
On the other hand, denoting with $M(n_{T})$ the measurement matrix
corresponding to a thermal state $\rho(n_{T})$, it is easy to
prove that $M(0)-M(n_{T})\geq0$ $\forall n_{T}\geq0$. From this
relation and Eq.~(\ref{gamma0}), one has that $F[\hat{U}^{\dagger
}\rho\hat{U},\rho(n_{T})]\leq
F[\hat{U}^{\dagger}\rho\hat{U},\rho(0)]$, but
$F^{(0)}=F[\hat{U}^{\dagger}\rho\hat{U},\rho(n_{T})]\leq
F[\hat{U}^{\dagger}\rho\hat{U},\rho(0)]=F[\rho,\hat{U}\rho(0)\hat{U}^{\dagger
}]$ with $\hat{U}\rho(0)\hat{U}^{\dagger}\equiv\hat{E}_{0,p}$ pure
Gaussian state.
\end{proof}

According to this latter result, the optimal Gaussian measurement
operator $\hat{E}_{0}$ is actually a projection onto a pure
Gaussian state and therefore has to be searched within the set of
squeezed states $\left|  \alpha,\varepsilon\right\rangle =\hat{D}
(\alpha)\hat{S}(\varepsilon)\left|  0\right\rangle $. Here
$\hat{D}(\alpha)$ is the displacement operator with $\alpha$
complex amplitude, while $\hat{S}(\varepsilon)$ is the squeezing
operator with $\varepsilon\equiv r\exp(2i\varphi)$ and $r,\varphi$
are the squeezing factor and phase respectively \cite{milburn}.
Since the CM of the input state, $V_{in}$, is given, Charlie has
to optimize the protocol only with respect to the CM $V_{0}$ of
the squeezed state $\left|  \alpha ,\varepsilon\right\rangle $,
which is given by
\begin{equation}
V_{0}(\xi,\varphi)=\frac{1}{2}\left(
\begin{array}
[c]{cc}
\xi\sin^{2}\varphi+\xi^{-1}\cos^{2}\varphi & (\xi-\xi^{-1})\cos\varphi
\sin\varphi\\
(\xi-\xi^{-1})\cos\varphi\sin\varphi & \xi\cos^{2}\varphi+\xi^{-1}\sin
^{2}\varphi
\end{array}
\right)  \label{V0}
\end{equation}
where $\xi\equiv\exp(2r)$, and therefore the optimization has to
be done with respect to $\xi$ and $\varphi$. Finding a global
maximum point $(\bar{\xi},\bar{\varphi})$ is difficult in general
and therefore we split the problem in two steps: we first maximize
the fidelity $F^{(0,p)}\equiv F(\xi,\varphi)$ with respect to
$\xi$\ for an arbitrary but fixed $\varphi$, and then we maximize
the result $F(\bar{\xi} (\varphi),\varphi)$ with respect to
$\varphi$. Mathematically speaking the function $F(\xi,\varphi)$
is bounded and continuous in the domain
$]0,+\infty\lbrack\times\lbrack0,\pi]$ and we implicitly have to
consider its continuous extension in
$[0,+\infty]\times\lbrack0,\pi]$ in order to surely have the
existence of global extremal points. The first step of
maximization is solved by the following
\begin{proposition}
For every squeezing phase $\varphi\in[0,\pi]$, Charlie can select
a squeezing factor $\bar{\xi}(\varphi)$ such that
$F(\bar{\xi}(\varphi),\varphi)\geq F(\xi,\varphi)$ $\forall\xi$
(phase-dependent global maximum point). The point
$\bar{\xi}(\varphi)$ can be derived analytically from the CMs $V$
and $V_{in}$, according to the following four-step procedure:
\begin{enumerate}
\item{Construct the matrices $\Gamma^{tr}$ of Eq.~(\ref{gammatr}), $\Sigma
$ of Eq.~(\ref{sigma}) and $U\equiv\Sigma\Omega\Gamma^{tr}\Omega^{T}\Sigma^{T}$.}
\item{Define a 2-D vector
\begin{equation}
\vec{u}\equiv\left(
\begin{array}
[c]{c}
\det C+1/4\\
(\det\Sigma)^{2}-{\rm Tr}(\Omega C\Omega^{T}U)
\end{array}
\right)  \label{u}
\end{equation}
and a 2-D phase-dependent vector
\begin{equation}
\vec{k}(\varphi)\equiv\left(
\begin{array}
[c]{c}
\vec{\varphi}^{T}U\vec{\varphi}\\
\vec{\varphi}^{T}C\vec{\varphi}
\end{array}
\right)  \label{kappa}
\end{equation}
where $\vec{\varphi}^{T}\equiv(\sin\varphi,\cos\varphi)$.}
\item{Define the scalar product
\begin{equation}\label{scalar}
\gamma(\varphi)\equiv\vec{u}\cdot\vec {k}(\varphi)
\end{equation} and the third component of the vector product
\begin{equation}\label{vector}
\omega(\varphi)\equiv\frac{1}{2}[\vec{k}(\varphi)\times\vec{k}(\varphi-\pi/2)]_{z}.
\end{equation}}
\item{Denote with $p(\varphi)$ the $\varphi$-dependent logic proposition
$\gamma(\varphi)<0\wedge\gamma(\varphi-\pi/2)<0$.}
\end{enumerate}
Then:
\begin{align}
p(\varphi) &  =1\Longleftrightarrow\bar{\xi}(\varphi)=\frac{\omega
(\varphi)-\sqrt{\omega(\varphi)^{2}+\gamma(\varphi-\pi/2
)\gamma(\varphi)}}{\gamma(\varphi-\pi/2)}\equiv\xi_{-}(\varphi)\label{sol1}\\
p(\varphi) &
=0\Longleftrightarrow\bar{\xi}(\varphi)=0\text{\quad}\vee\quad
\bar{\xi}(\varphi)=+\infty.\label{sol2}
\end{align}
\end{proposition}
See Appendix B for the proof.

The second step concerns the \textit{maximization over the
squeezing phase} $\varphi$. Using vectors
$\vec{u}^{T}=(u_{x},u_{y})$ in (\ref{u}) and $\vec
{k}(\varphi)^{T}=(k_{x}(\varphi),k_{y}(\varphi))$ in
(\ref{kappa}),\ the fidelity can be written as
\begin{equation}\label{Fvector}
F(\xi,\varphi)=\left[
\det\Gamma^{tr}-\frac{-u_{y}+(\xi/2)k_{x}(\varphi
-\pi/2)+(\xi^{-1}/2)k_{x}(\varphi)}{u_{x}+(\xi/2)k_{y}(\varphi-\pi
/2)+(\xi^{-1}/2)k_{y}(\varphi)}\right]  ^{-1/2} .
\end{equation}
We then consider the piecewise continuous function of $\varphi$,
$\xi=\bar {\xi}(\varphi)$ defined according to (\ref{sol1}),
(\ref{sol2})\ and the corresponding phase-dependent teleportation
fidelity $\bar{F}(\varphi)\equiv F(\bar{\xi }(\varphi),\varphi)$
which is continuous on $[0,\pi]$. From Eq.~(\ref{V0}) one has
\begin{equation}
V_{0}(\xi,\varphi)=V_{0}(\xi,\varphi+\pi)=V_{0}(\xi^{-1},\varphi
+\pi/2)\label{periodicity}
\end{equation}
and therefore $F(0,\varphi)=F(+\infty,\varphi+\pi/2)$ and
$F(0,\varphi +\pi/2)=F(+\infty,\varphi)$. This implies that
finding the maximum point $\bar{\varphi}$ of $\bar{F}(\varphi)$ is
equivalent to find the maximum point of the piecewise continuous
function
\begin{equation}
\tilde{F}(\varphi)=\left\{
\begin{array}
[c]{c}
F(\xi_{-}(\varphi),\varphi)\qquad\quad\qquad\qquad\qquad\qquad\qquad\mathrm{if}
\quad p(\varphi)=1\quad\\
F(0,\varphi)=[\det\Gamma^{tr}- k_{x}(\varphi)/k_{y}(\varphi)
]^{-1/2}\qquad\mathrm{if}\quad p(\varphi)=0\quad.
\end{array}
\right.  \label{Ftilde}
\end{equation}
Now, one has three cases: i) $\bar{\varphi}$ is a stationary point
of $F(0,\varphi)$; ii) $\bar{\varphi}$ is a stationary point of
$F(\xi _{-}(\varphi),\varphi)$; iii) $\bar{\varphi}$ is one of the
border points dividing the intervals where $p(\varphi)=1$ from
those where $p(\varphi)=0$. We report a simple analytical
expression of the final global maximum point only in the first
case, while in the other two cases the expressions are extremely
involved. In case i), defining the $2\times2$\ matrix $\tau\equiv
U\Omega C\Omega^{T}$, the stationary points $\varphi_{\pm}$ of
$F(0,\varphi)$ are given by the relation
\begin{equation}
\cos2\varphi_{\pm}=\frac{\tau_{12}^{2}-\tau_{21}^{2}\pm(\tau_{11}-\tau
_{22})\sqrt{(\tau_{11}-\tau_{22})^{2}+4\tau_{12}\tau_{21}}}{(\tau_{11}
-\tau_{22})^{2}+(\tau_{12}+\tau_{21})^{2}}.\label{FiMax}
\end{equation}
In many cases of practical interest (for instance when coherent
states or $\varphi=0$ squeezed states are teleported through a CM
$V$ with diagonal blocks, as for example in
\cite{Braun,PRL,PRA,parisopto}), the above procedure allows to
find the maximum point $(\bar{\xi}(\bar{\varphi}),\bar{\varphi})$
and the corresponding optimal conditional fidelity
$F^{(0)}_{max}=F(\bar{\xi}(\bar{\varphi}),\bar{\varphi})$ quite
quickly. In some easy cases when matrices $U$ and $C$ are
proportional to the identity, we see from (\ref{kappa}) that the
above optimization becomes $\varphi$-independent and therefore the
maximum point is given by $\bar{\xi}=1$ if $\gamma<0$ or by
$\bar{\xi}=0$ if $\gamma\geq0$. In the first case the optimal
Gaussian state is a coherent state, i.e. $\hat{E}_{0}^{opt}=\left|
\alpha\right\rangle \left\langle \alpha\right|  $ (with $\alpha$
arbitrary), while in the second case it is an infinitely squeezed
state, i.e. $\hat{E}_{0}^{opt}=\left| X(\varphi )\right\rangle
\left\langle
X(\varphi )\right| $ where $\hat{X}(\varphi )\equiv 2^{-1/2}(\hat{c}%
\,e^{-i\varphi }+\hat{c}^{\dagger }\,e^{i\varphi })$ (phase
$\varphi $ and eigenvalue $X(\varphi)$ are arbitrary).

\section{Local Gaussian measurement}

The above optimization results (propositions 2 and 3) refer to the
conditional scheme where Charlie performs a dichotomic measurement
and the Gaussian outcome $n=0$ is selected, and they can be
directly used for the conditional generation of entanglement in
that configuration. However, these results can be extended to a
different scheme where all the measurement outcomes are Gaussian
and therefore all the conditional fidelities can potentially
outperform $F^{tr}$ according to Proposition~1. More in detail,
Proposition~1 suggests to consider for Charlie a measurement which
creates a conditional bipartite state which is Gaussian for every
outcome, and we surely achieve this condition if we consider for
Charlie a local Gaussian measurement, i. e., a local measurement
$\{\hat{E}(n)\}$ transforming a Gaussian multipartite state into
another Gaussian state for every measurement outcome $n$. Notice
that here we include in the Gaussian states also \emph{asymptotic}
Gaussian states, such as the infinitely squeezed states. Examples
of local Gaussian measurements are provided by heterodyne
measurement on a single mode $c$, i.e.,
$\{|\alpha\rangle\langle\alpha|/\sqrt{\pi},\alpha
\in\mathbb{C}\}$, or they are obtained when the $c$ mode is
coupled to ancillary modes by a Gaussian unitary interaction and
then the ancillas are subject to heterodyne or homodyne
measurement. Consider then an assisted protocol where Charlie
performs a local Gaussian measurement $\{\hat{E}(n)\}$ on his mode
$c$ and classically communicates the measurement result $n$ to Bob
(who, in turn, makes a drift-cancelling displacement depending
upon the measurement outcomes of both Alice and Charlie). It is
possible to prove a result analogous to Proposition~2 of the
dichotomic case:

\begin{proposition}
For every local Gaussian measurement $\{\hat{E}(n)\}$ with
fidelity $F$, there exists a ``pure'' local Gaussian measurement
$\{\hat{E}_{\varepsilon}(\alpha)\equiv\left| \alpha,\varepsilon
\right\rangle \left\langle \alpha,\varepsilon\right|
/\sqrt{\pi},\alpha \in\mathbb{C}\}$ with a suitable $\varepsilon$,
such that its fidelity $F(\varepsilon)\geq F$.
\end{proposition}

\begin{proof}
Suppose that Charlie performs an arbitrary local Gaussian
measurement $\{\hat{E}(n)\}$\ on his mode $c$, so that the
conditional reduced state of Alice and Bob is given by
$\rho^{(n)}=P(n)^{-1}\mathrm{Tr}_{c}(\hat{E}(n)\rho
\hat{E}(n)^{\dagger})$ corresponding to a fidelity $F^{(n)}$ (the
effective fidelity of the protocol is the average over the results
$F=\sum _{n}P(n)F^{(n)}$). Suppose now that Charlie performs a
further local measurement on $c$\ given by a dichotomic
measurement and the Gaussian outcome $\hat{E}_{0}$ has been
selected. In such a case the reduced state will be
$\rho^{(n)\prime}\propto\mathrm{Tr}_{c}(\hat{E}_{0}\hat{E}(n)\rho
\hat{E}(n)^{\dagger}\hat{E}_{0}^{\dagger})=\mathrm{Tr}_{c}(\hat{E}(n)^{\dagger
}\hat{E}_{0}^{2}\hat{E}(n)\rho)=\mathrm{Tr}_{c}(\hat{E}(n)^{\prime}
\rho\hat{E}(n)^{\prime\dagger})$ where $\hat{E}(n)^{\prime}\equiv
\lbrack\hat{E}(n)^{\dagger}\hat{E}_{0}^{2}\hat{E}(n)]^{1/2}$ is a
Gaussian state \cite{distill}, and the corresponding fidelity will
be $F^{(n)\prime}$. From Eq.~(\ref{result1}) we have $F^{(n)}\leq
F^{(n)\prime}$ $\forall n$, implying
$F\leq\sum_{n}P(n)F^{(n)\prime}\leq F^{(\tilde{n})\prime}$ where
$F^{(\tilde{n})\prime}$ is the maximum value achieved for a
particular outcome $\tilde{n}$. Now, Proposition~2 tells us that
there exists a pure Gaussian state $\hat{E}_{p}\equiv\left|
\alpha,\varepsilon\right\rangle \left\langle
\alpha,\varepsilon\right|  $ such that $F^{(p)}\geq
F^{(\tilde{n})\prime}$, with a suitably chosen squeezing complex
factor $\varepsilon$, while $\alpha$ can be arbitrary. It is then
evident that one has a teleportation fidelity $F^{(p)}$ also if
Charlie directly applies the measurement
$\hat{E}_{\varepsilon}(\alpha)\equiv\left|  \alpha,\varepsilon
\right\rangle \left\langle \alpha,\varepsilon\right| /\sqrt{\pi}$
on the tripartite state $\rho$ and classically communicates the
result $\alpha$ to Bob, so that one has
$F(\varepsilon)=F^{(p)}\geq F^{(\tilde{n})\prime}\geq F$.
\end{proof}

Trivially the previous proposition assures the existence of a
local Gaussian measurement of the pure form
$\{\hat{E}_{\varepsilon}(\alpha)\equiv\left|
\alpha,\varepsilon\right\rangle \left\langle
\alpha,\varepsilon\right| /\sqrt{\pi},\alpha\in\mathbb{C}\}$ which
leads to an assisted fidelity $F(\varepsilon)\geq F^{tr}$. In
fact\ it is sufficient to consider $\{\hat{E}(n)\}=I$ and apply
the proposition. More importantly it implies that the optimal
local Gaussian measurement must be searched for within the set of
pure measurements
$\{\{\hat{E}_{\varepsilon}(\alpha)\},\varepsilon\in\mathbb{C}\}$,
which is equivalent to maximize with respect to the $2\times2$ CM
$V_{0}(\varepsilon )=V_{0}(\xi,\varphi)$ of Eq.~(\ref{V0}). Thanks
to this result, the optimization procedure is exactly the one
given for the dichotomic measurement, i.e., it is given by the
maximization over the squeezing factor as in Proposition~3 and by
the subsequent maximization over the squeezing phase. Repeating
such procedure it is possible to find an optimal pair of
parameters $(\bar{\xi}(\bar{\varphi}),\bar{\varphi})$ which
describes the optimal local Gaussian measurement
$\{\hat{E}_{\bar{\varepsilon}}(\alpha)\}$ and provides the
corresponding optimal assisted fidelity
$F(\bar{\xi},\bar{\varphi})$ using (\ref{Fvector}). Notice that
for finite squeezing ($\bar{\xi}\neq0,+\infty$) the measurement
$\{\hat{E}_{\bar {\varepsilon}}(\alpha)\}$ can be realized by
first applying a unitary squeezing transformation
$\hat{S}(\bar{\varepsilon})$ to mode $c$\ and then making
heterodyne detection. For infinite squeezing
($\bar{\xi}=0,+\infty$) the measurement
$\{\hat{E}_{\bar{\varepsilon}}(\alpha)\}$ is instead equivalent to
a homodyne detection, i.e., to $\left|
X(\bar{\varphi})\right\rangle _{c}\left\langle
X(\bar{\varphi})\right| $ for $\bar{\xi}=+\infty$ and to $\left|
X(\bar{\varphi}+\pi/2)\right\rangle _{c}\left\langle
X(\bar{\varphi}+\pi/2)\right| $ for $\bar{\xi}=0$, where
$\hat{X}(\varphi
)\equiv2^{-1/2}(\hat{c}\,e^{-i\varphi}+\hat{c}^{\dagger}\,e^{i\varphi})$.
As for the dichotomic case, we can use this optimized measurement
to create conditional bipartite entanglement between Alice and Bob
and now this can be done in a deterministic way since all the
outcomes are Gaussian. Note that this is not in contrast with the
impossibility of entanglement distillation of Gaussian states with
local Gaussian operations and classical communications
\cite{distill} because here we only have a transfer of
entanglement resources from a tripartite to a bipartite state. In
the following section we give explicit examples of application of
our optimization procedure.

\section{Examples}
\subsection{Optimization of fidelity}
As an example of application of our theoretical results, we
consider a three-mode Gaussian state with CM
\begin{equation}
V=\left(
\begin{array}
[c]{ccc}
qI & wR & wR\\
wR & sI & tI\\
wR & tI & sI
\end{array}
\right)  \label{CMgame}
\end{equation}
where $ R={\rm diag}\left(1,-1\right)$, $I$ is the $2\times2$
matrix identity, and the coefficients $q,s,t$ and$\ w$ are real
numbers. The CM (\ref{CMgame}) represents a \textit{genuine} CM
(i.e. it corresponds to a physical state) if and only if
\cite{Cirac3modes}
\begin{equation}
V-\frac{i}{2}J\geq0 \label{genuine}
\end{equation}
where
\begin{equation}
J\equiv\left(
\begin{array}
[c]{ccc}
\Omega &  & \\
& \Omega & \\
&  & \Omega
\end{array}
\right)
\end{equation}
with $\Omega$ given in (\ref{Omlabel}). Setting $s=(q+1)/2,$
$t=q/2$ and $w=[\sqrt{(2q-1)(q+1)}]/2$ in (\ref{CMgame}), we have
a $q-$dependent\textit{ }CM $V(q)$ which is genuine for every
$q\geq1/2$. This is exactly the correlation matrix considered in
\cite{game}, where a novel (cooperative) telecloning protocol is
proved to outperform the standard (non cooperative) one for
increasing values of $q$. We call $q$ the `noise parameter' since
it determines the linear entropy of the bipartite reduced state
$\rho_{tr}\equiv Tr_{c}(\rho)$ used by Alice and
Bob in the non-assisted protocol, i.e., $S(\rho_{tr})=1-Tr_{ab}(\rho_{tr}%
^{2})=q/(q+1)$.

\begin{figure}[ptbh]
\vspace{-0.0cm}
\par
\begin{center}
\includegraphics[width=0.35\textwidth]{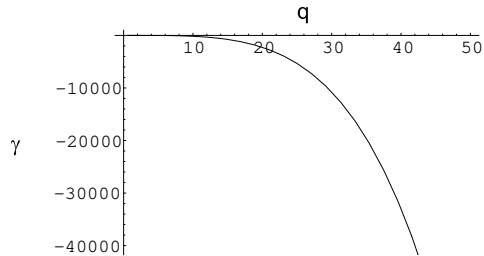}
\end{center}
\par
\vspace{-0.0cm}\caption{$\varphi$-independent scalar product
$\gamma$ versus the noise parameter $q$ for $1/2\leq q\leq50$.}
\label{gammagame}
\end{figure}

\begin{figure}[ptbh]
\vspace{-0.0cm}
\par
\begin{center}
\includegraphics[width=0.35\textwidth]{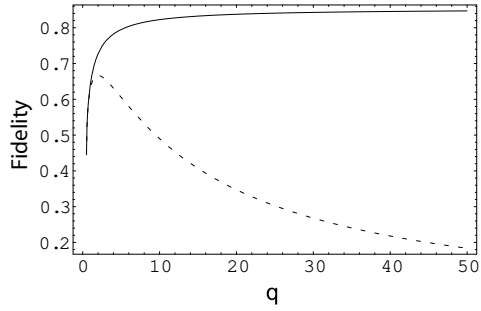}
\end{center}
\par
\vspace{-0.0cm}\caption{Optimal assisted fidelity $F(q)$ (solid
line) and non-assisted fidelity $F^{tr}(q)$ (dashed line) versus
the noise parameter $q$ for $1/2\leq q\leq50$.} \label{fidelgame}
\end{figure}

Suppose that Alice, Bob and Charlie possess modes ${\hat{a}}$,
$\hat{b}$ and ${\hat{c}}$ respectively, and Alice wants to
teleport a coherent state ($V_{in}=I/2$) to Bob with the help of
Charlie. Due to the simple form of $V$ and $V_{in}$, it is
straightforward to compute the optimal local Gaussian measurement
which Charlie can perform in order to optimize teleportation
fidelity. From Proposition~4 we know that it has the pure form
$\{\left| \alpha ,\varepsilon\right\rangle \left\langle
\alpha,\varepsilon\right| /\sqrt{\pi },\alpha\in\mathbb{C}\}$ and
applying Proposition~3 we can calculate the value
$\bar{\varepsilon}\equiv(\bar{\xi},\bar{\varphi})$ corresponding
to a maximum. The procedure of Proposition~3 goes as follows:
\begin{enumerate}
\item $\Gamma^{tr}=hI$ \ with $\ h\equiv(1+q+s-2w),$

$\Sigma=(w-t)I,$

$U=(w-t)^{2}hI.$

\item $\vec{u}=\left(
\begin{array}
[c]{c}
s^{2}+1/4\\
(w-t)^{2}[(w-t)^{2}-2sh]
\end{array}
\right)  ,$

$\vec{k}(\varphi)=\left(
\begin{array}
[c]{c}
(w-t)^{2}h\\
s
\end{array}
\right)  \equiv\vec{k}.$

Note that, since Charlie's submatrix $sI$ and matrix $U$ are
proportional to the identity, vector $\vec{k}(\varphi)$ becomes
$\varphi-$independent and therefore all the subsequent
optimization procedure becomes $\varphi-$independent.

\item $\gamma=(w-t)^{2}[s(w-t)^{2}-s^{2}h+h/4],$

$\omega=0.$

\item  As already pointed out at the end of section~III, we have the following
simplification for Charlie's optimal measurement
\begin{align}
\gamma &  <0\Longleftrightarrow\bar{\xi}=1\\
\gamma &
\geq0\Longleftrightarrow\bar{\xi}=0.
\end{align}
The first case corresponds to an heterodyne detection while the
second case corresponds to an homodyne detection.
\end{enumerate}
We have $\gamma<0$ for every $q\geq1/2$ (see Fig.~\ref{gammagame})
and therefore heterodyne detection ($\bar{\xi}=1$,
$\bar{\varphi}$~arbitrary)\ is the optimal local Gaussian
measurement which Charlie can perform in order to maximize
teleportation fidelity of coherent states between Alice and Bob
with this kind of shared channel. The corresponding optimal
assisted fidelity $F(\bar{\xi},\bar{\varphi})$ is computed from
(\ref{Fvector}) setting $\bar{\xi}=1$ and $\bar{\varphi}$
arbitrary, and it is given by
\begin{equation}
F=\{h^{2}-(w-t)^{2}(s+1/2)^{-2} [(2s+1)h-(w-t)^{2}]\}^{-1/2}
\end{equation}
which is a function of the noise parameter $q$.
Fig.~\ref{fidelgame} clearly shows the improvement provided by the
optimal assisted fidelity $F(q)$ with respect to the non-assisted
fidelity $F^{tr}(q)=h^{-1}$ for every $q$ and especially for
increasing noise in the channel.

\subsection{Conditional generation of entanglement}

Consider now a three-mode Gaussian state $\rho$ with CM
\begin{equation}
V=\left(
\begin{array}
[c]{ccc}
aI & fI & eR\\
fI & bI & dR\\
eR & dR & cI
\end{array}
\right)  \label{CMnumber}
\end{equation}
where $ R={\rm diag}\left(1,-1\right)$, $I$ is the $2\times2$
matrix identity and
\begin{equation}
\{a,b,c,d,e,f\}=\{10.15,5.52,15.2,8.87,12.3,6.96\}. \label{param1}
\end{equation} Numerical values in (\ref{param1}) are chosen so that CM
(\ref{CMnumber}) represents a genuine CM. Tracing out mode
${\hat{c}}$ (Charlie), the remaining bipartite state
$\rho^{tr}=Tr_{c}(\rho)$ of modes ${\hat{a}}$ and $\hat{b}$ has CM
\begin{equation}
V^{tr}=\left(
\begin{array}
[c]{cc}
aI & fI\\
fI & bI
\end{array}
\right)
\end{equation}
and one can verify the separability condition \cite{Simon}
\begin{equation}
V^{tr}-\frac{i}{2}\tilde{J}\geq0 \label{SepCond}
\end{equation}
where
\begin{equation}
\tilde{J}\equiv\left(
\begin{array}
[c]{cc}
\Omega & \\
& -\Omega
\end{array}
\right).
\end{equation}
Condition (\ref{SepCond}) means that the reduced state
$\rho^{tr}$\ shared by Alice and Bob is a separable state and
therefore it cannot allow a \textit{quantum} teleportation, i. e.
it leads to\ a fidelity $F^{tr}\leq1/2$ for teleportation of
coherent states.

\begin{figure}[ptbh] \vspace{-0.0cm}
\par
\begin{center}
\includegraphics[width=0.35\textwidth]{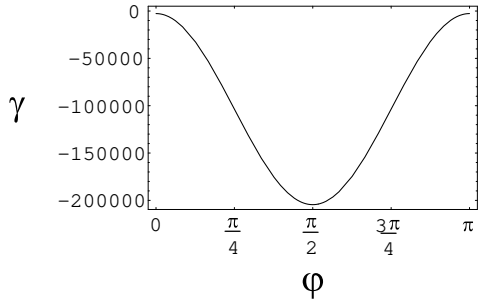}
\end{center}
\par
\vspace{-0.0cm}\caption{Scalar product $\gamma$ versus the
squeezing phase $\varphi$. This refers to a tripartite Gaussian
state having CM (\ref{CMnumber}) with parameter choice
(\ref{param1}).} \label{gamma2}
\end{figure}

\begin{figure}[ptbh]
\vspace{-0.0cm}
\par
\begin{center}
\includegraphics[width=0.35\textwidth]{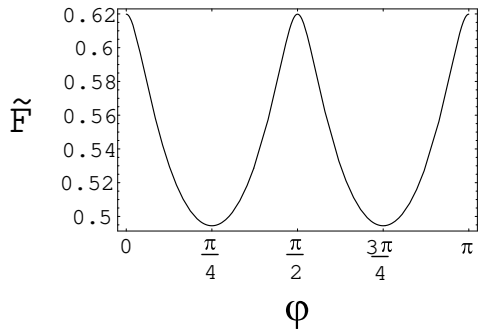}
\end{center}
\par
\vspace{-0.0cm}\caption{Fidelity $\tilde{F}$ versus the squeezing
phase $\varphi$. This refers to a tripartite Gaussian state having
CM (\ref{CMnumber}) with parameter choice (\ref{param1}).}
\label{fidel2}
\end{figure}

Here we give an explicit example where Charlie can conditionally
create bipartite entanglement between Alice and Bob by performing
an optimal Gaussian measurement at his site and then communicating
the result. Consider, for simplicity, teleportation of coherent
states, and compute the optimal local Gaussian measurement
applying the procedure of Proposition~3:

\begin{enumerate}
\item $\Gamma^{tr}=(a+b+1)I-2fR,$

$\Sigma=eI-dR,$

$U=\left(
\begin{array}
[c]{cc}
(e-d)^{2}(a+b+1+2f) & 0\\
0 & (e+d)^{2}(a+b+1-2f)
\end{array}
\right)  .$

\item $\vec{u}=\left(
\begin{array}
[c]{c}
c^{2}+1/4\\
(e^{2}-d^{2})^{2}-2c\Lambda_{0}
\end{array}
\right)  ,$

$\vec{k}(\varphi)=\left(
\begin{array}
[c]{c}
\Lambda_{0}+\Lambda_{1}\cos(2\varphi)\\
c
\end{array}
\right)  ,$

where $\Lambda_{0}\equiv(e^{2}+d^{2})(a+b+1)-4def$ \ and
$\Lambda_{1} \equiv-2f(e^{2}+d^{2})+2de(a+b+1).$

\item $\gamma(\varphi)=c(e^{2}-d^{2})^{2}-(c^{2}-1/4)\Lambda_{0}
+(c^{2}+1/4)\Lambda_{1}\cos(2\varphi),$

$\omega(\varphi)=c\Lambda_{1}\cos(2\varphi).$
\end{enumerate}
The next step concerns the study of the value of the proposition
$p(\varphi ):\gamma(\varphi)<0\wedge\gamma(\varphi-\pi/2)<0$.
Fig.~\ref{gamma2} shows that $\gamma(\varphi)<0$ for every
$\varphi$,$\ $and therefore$\ p(\varphi)=1$ for every $\varphi$,
so that the phase-dependent global maximum point $\bar{\xi
}(\varphi)$\ is always given by $\xi_{-}(\varphi)$ as in
(\ref{sol1}). For this reason the maximization over the squeezing
phase $\varphi$\ is equivalent to find the maximum point of
$\tilde{F}(\varphi)=F(\xi_{-}(\varphi),\varphi)$ (see
(\ref{Ftilde}) and (\ref{Fvector} )), which we have plotted in
Fig.~\ref{fidel2}. Maximum points take the values
$\varphi_{k}=k\pi/2,$ $k=0,1,...$ so that we can choose
$\bar{\varphi}=0$, which gives $\bar{\xi
}=\bar{\xi}(\bar{\varphi})\sim0.087$ and
$F(\bar{\xi},\bar{\varphi})=\tilde {F}(\bar{\varphi})\sim62\%$. In
conclusion, Charlie's optimal local Gaussian measurement is
equivalent to a squeezing transformation
$\hat{S}(\bar{\varepsilon})$\ of his mode given by
$\bar{\varepsilon}\equiv(\bar{\xi},\bar{\varphi})=(0.087,0)$
followed by an heterodyne detection. Such a measurement (and the
subsequent classical communication) implies a fidelity of $62\%$
for the teleportation of coherent states and therefore it is
sufficient to create bipartite entanglement between Alice and Bob.

\subsubsection{Different parameter choice} 
\begin{figure}[ptbh]
\vspace{-0.0cm}
\par
\begin{center}
\includegraphics[width=0.6\textwidth]{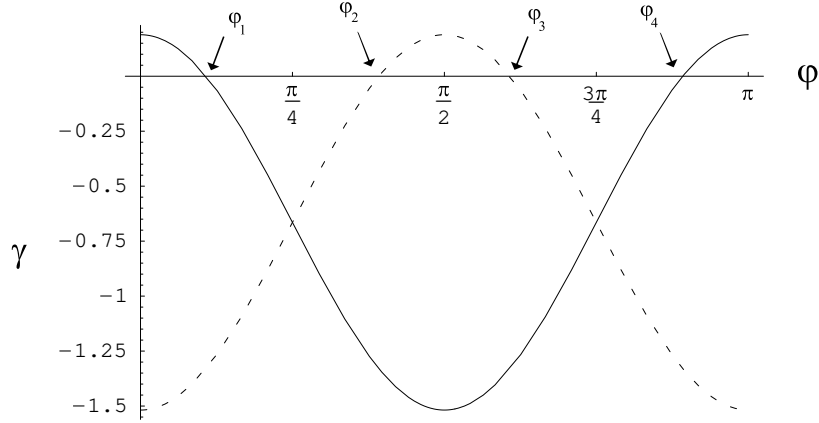}
\end{center}
\par
\vspace{-0.0cm}\caption{Scalar product $\gamma(\varphi)$ (solid
line) and its displaced form $\gamma(\varphi-\pi/2)$ (dashed line)
versus the squeezing phase $\varphi$. Denoting with $\varphi_{k}$
($k=1,2,3,4$) their intersections with the $\varphi$-axis, the
proposition $p(\varphi):$ $\gamma(\varphi )<0\wedge$
$\gamma(\varphi-\pi/2)<0$ is true only inside the region
$\mathcal{R}
\equiv]\varphi_{1},\varphi_{2}[\cup]\varphi_{3},\varphi_{4}[ $.
This refers to a tripartite Gaussian state having CM
(\ref{CMnumber}) with parameter choice (\ref{param2}).}
\label{gamma3}
\end{figure}

\begin{figure}[ptbh]
\vspace{-0.0cm}
\par
\begin{center}
\includegraphics[width=0.6\textwidth]{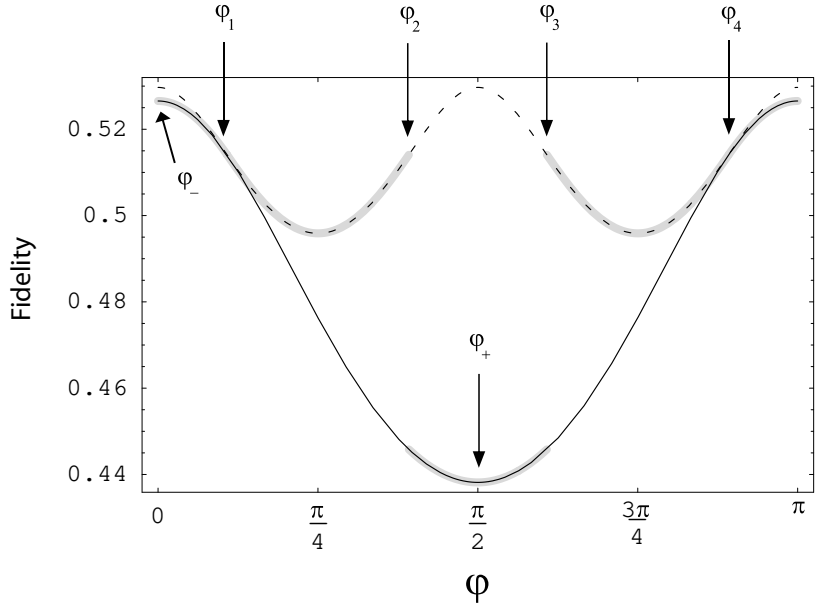}
\end{center}
\par
\vspace{-0.0cm}\caption{Fidelities $F(\xi_{-}(\varphi),\varphi)$
(dashed line), $F(0,\varphi)$ (solid line) and
$\tilde{F}(\varphi)$ (thicker line) versus the squeezing phase
$\varphi$. The piecewise continuous function $\tilde{F}(\varphi)$
is discontinuous at points $\varphi_{2}$ and $\varphi _{3}$. This
refers to a tripartite Gaussian state having CM (\ref{CMnumber})
with parameter choice (\ref{param2}).} \label{fidel3}
\end{figure}

It is instructive to
study a different parameter choice leading to a more involved
situation. Setting
\begin{equation}
\{a,b,c,d,e,f\}=\{0.55,0.89,0.94,0.74,0.249,0.12\}\label{param2}
\end{equation}
in (\ref{CMnumber}), we have again a genuine CM and a separable
reduced state $\rho^{tr}$ for Alice and Bob. However in this case
the proposition $p(\varphi)$ is true only for
$\varphi\in]\varphi_{1},\varphi_{2} [\ \cup\
]\varphi_{3},\varphi_{4}[\equiv\mathcal{R}$, where the border
points $\varphi_{k}$\ are given by $\varphi_{1}=0.339,$
$\varphi_{2}=\pi/2-0.339,$ $\varphi_{3}=\pi/2+0.339,$ and
$\varphi_{4}=\pi-0.339$ (see Fig.~\ref{gamma3}). Maximization over
the squeezing phase is given by the maximization of the piecewise
continuous function
\begin{equation}
\tilde{F}(\varphi)=\left\{
\begin{array}
[c]{l}
F(\xi_{-}(\varphi),\varphi)\qquad\qquad\qquad\qquad\qquad\qquad\quad
\ \mathrm{if}\quad\varphi\in\mathcal{R}\\
F(0,\varphi)=[\det\Gamma^{tr}-k_{x}(\varphi)/k_{y}(\varphi)]^{-1/2}
\qquad\mathrm{if}\quad\varphi\in\lbrack0,\pi\lbrack-\mathcal{R}
\end{array}
\right.
\end{equation}
(see (\ref{Ftilde}) and (\ref{Fvector})). In Fig.~\ref{fidel3} we
see that $F(\xi_{-}(\varphi),\varphi)$ does not have maximum
points inside the region $\mathcal{R}$, while $F(0,\varphi)$ has
two stationary points $\varphi_{-}=0$ and $\varphi_{+}=\pi/2$
which fall in $[0,\pi\lbrack-\mathcal{R}$. They are exactly the
ones derived from (\ref{FiMax}) as we can easily check noting that
$\tau_{12}=\tau_{21}=0$ and $\tau_{11}-\tau_{22}<0$ so that $\cos
(2\varphi_{\pm})=\mp1$. It is evident from Fig.~\ref{fidel3} that
$\varphi_{-}$ is a maximum point while $\varphi_{+}$ is a minimum
point for $F(0,\varphi)$. In order to find the global maximum, we
have to compare $\tilde{F}(\varphi _{-})=F(0,\varphi_{-})$ with
the right and left limits of $\tilde{F}(\varphi)$ at the border
points $\varphi_{k}$. Since $\tilde{F}(\varphi_{1})=\tilde
{F}(\varphi_{4})=F(\xi_{-}(\varphi_{2}),\varphi_{2})=F(\xi_{-}(\varphi
_{3}),\varphi_{3})=0.514,$
$F(0,\varphi_{2})=F(0,\varphi_{3})=0.446$, and
$F(0,\varphi_{-})=0.526$, we have that
$\bar{\varphi}=\varphi_{-}=0$. In conclusion, in this different
parameter choice, the optimal Gaussian measurement at Charlie's
site is the homodyne detection $\left| X(\pi/2)\right\rangle
_{c}\left\langle X(\pi/2)\right|  $ which implies a fidelity of
$52.6\%$ for teleportation of coherent states, and therefore it is
again sufficient (together with the classical communication of the
result) to create bipartite entanglement between Alice and Bob.

\section{Conclusion}

We have studied how a two-party teleportation process (between
Alice and Bob) within a three-party shared quantum channel can be
conditioned by a local measurement and a classical communication
of the third party (Charlie). In particular our analysis has been
carried out for a shared Gaussian channel, the teleportation of
pure Gaussian states, and two general kinds of local measurement
at Charlie's site. We have first shown the case of a dichotomic
measurement and we have proved that the non-Gaussian outcome
always worsens the fidelity while the Gaussian outcome always
improves it, even allowing the conditional generation of
entanglement. Then we have shown how the dichotomic measurement
can be designed so that the Gaussian outcome optimizes the
teleportation fidelity and we have extended such results directly
to the case of a local Gaussian measurement at Charlie's site.
From the knowledge of the correlation matrices (the one of the
shared tripartite Gaussian state and the one of the state to be
teleported), Charlie can always determine and perform an optimal
local Gaussian measurement given by a set of squeezed states with
squeezing factor $\bar{\xi}(\bar{\varphi})$ and squeezing phase
$\bar{\varphi}$, maximizing the fidelity of teleportation of pure
Gaussian states between Alice and Bob.

It is an interesting and still open question to establish if this
optimal Gaussian measurement is also the best among \emph{all}
possible measurements at Charlie's site. Proposition~1 shows that
in the dichotomic case, the Gaussian outcome yields always a
better result than the non-Gaussian one. This fact and also the
fact that we are here considering the particular task of
teleporting a one mode \emph{Gaussian} state employing a
tripartite \emph{Gaussian} state suggest that this optimal
Gaussian measurement can actually be the best possible measurement
Charlie can do to maximize this specific teleportation fidelity.
Notice that the recent paper of Ref.~\cite{Werner} has shown that,
for $1\rightarrow2$ cloning of coherent states, even though the
joint fidelity is maximized by a Gaussian cloner, the single-copy
fidelity is maximized by a \emph{non-Gaussian} cloner. However,
from the point of view of teleportation, the Ref.~[21] gives a
support to our conjecture that the fidelity of teleportation, for
Gaussian input and Gaussian quantum channel, is optimized by a
Gaussian measurement. In fact, in Ref.~[21], the particular
optimal non-Gaussian cloner which realizes $F_{1}=1$ and $F_{2}=0$
(with $F_{k}$ the single-copy fidelity of the $k^{th}$ clone), and
therefore gives the optimal quantum teleportation from the
coherent input to one clone, actually \emph{coincides} with a
Gaussian cloner (see Fig.$~$1 in Ref.~[21] at the extremal points
$(1,0)$ and $(0,1)$).

We finally notice that the procedure sketched in this work can be
applied to all CV teleportation networks based on a multipartite
Gaussian state and can be used also for the conditional generation
of bipartite entanglement.

\section{Acknowledgments}
Authors thank Jens Eisert for insightful comments.

\appendix

\section{Derivation of the correlation matrix of Eq.~(\ref{CM0})}

We can derive the CM of Eq.~(\ref{CM0}) using the symmetrically
ordered characteristic function \cite{milburn}. The measurement
operator $\hat
{E}_{0}$\ is a state of mode $c$ and it can be expressed as%
\begin{equation}
\hat{E}_{0}=\pi^{-1}\int\hat{D}_{c}^{\dagger}(\eta_{c})\Phi_{0}(\eta_{c}%
)d^{2}\eta_{c} \label{State_E0}%
\end{equation}
where $\hat{D}_{c}(\eta_{c})=\exp(\eta_{c}\hat{c}^{\dagger}-\eta_{c}^{\ast
}\hat{c})$ is the displacement operator acting on the Hilbert space of mode
$c$, $\Phi_{0}(\eta_{c})$ is the corresponding characteristic function, and
$\eta_{c}\equiv\eta_{c}^{R}+i\eta_{c}^{I}$ is a complex variable corresponding
to the annihilation operator $\hat{c}$ \cite{milburn}. Since $\hat{E}_{0}$ is
a Gaussian state, we have
\begin{equation}
\Phi_{0}(\vec{\eta}_{c})=\exp(-\vec{\eta}_{c}^{T}V_{0}\vec{\eta}_{c}+i\vec
{d}_{0}^{T}\vec{\eta}_{c}) \label{ChF_E0}
\end{equation}
where $V_{0}$ is the CM of the state, $\vec{d}_{0}\in{\mathbb{R}}^{2}$ the
displacement, and $\vec{\eta}_{c}^{T}\equiv(\eta_{c}^{I},-\eta_{c}^{R})$ an
${\mathbb{R}}^{2}$ vector connected to $\eta_{c}$. In the same way, the total
three-mode Gaussian state
\begin{equation}
\rho=\pi^{-3}\int\int\int\hat{D}_{a}^{\dagger}(\eta_{a})\hat{D}_{b}^{\dagger
}(\eta_{b})\hat{D}_{c}^{\dagger}(\eta_{c})\Phi(\eta_{a},\eta_{b},\eta
_{c})d^{2}\eta_{a}d^{2}\eta_{b}d^{2}\eta_{c} \label{State_tot}
\end{equation}
is associated to the characteristic function
\begin{equation}
\Phi(\vec{\eta})=\exp(-\vec{\eta}^{T}V\vec{\eta}+i\vec{d}^{T}\vec{\eta})
\label{ChF_tot}%
\end{equation}
where $V$ is the CM of Eq.~(\ref{CMtot}), $\vec{d}\in{\mathbb{R}}^{6}$ is the displacement, and
$\vec{\eta}^{T}\equiv(\vec{\eta}_{a}^{T},\vec{\eta}_{b}^{T},\vec{\eta}_{c}%
^{T})$ an ${\mathbb{R}}^{6}$ vector connected to $(\eta_{a},\eta_{b},\eta
_{c})$ as shown before.

The conditional reduced state
\begin{equation}
\rho^{(0)}\equiv P_{0}^{-1}\mathrm{Tr}_{c}(\hat{E}_{0}\rho\hat{E}_{0}%
^{\dagger})\label{State_red}%
\end{equation}
corresponds to a characteristic function $\Phi^{(0)}(\eta_{a},\eta_{b})$ which
is (by definition)
\begin{equation}
\Phi^{(0)}(\eta_{a},\eta_{b})\equiv\mathrm{Tr}_{ab}\left[  \rho^{(0)}\hat
{D}_{a}(\eta_{a})\hat{D}_{b}(\eta_{b})\right]  \label{ChF_red}%
\end{equation}
Putting Eqs.~(\ref{State_E0}) and (\ref{State_tot}) into Eq.~(\ref{State_red}), and the
subsequent result in Eq.~(\ref{ChF_red}), we obtain after some algebra
\begin{equation}
\Phi^{(0)}(\eta_{a},\eta_{b})=\pi^{-2}P_{0}^{-1}\int\int\Phi_{0}%
(\vartheta)\Phi_{0}(\kappa)\Phi(\eta_{a},\eta_{b},-\vartheta-\kappa
)\exp\left(  \frac{\vartheta\kappa^{\ast}-\vartheta^{\ast}\kappa}{2}\right)
d^{2}\vartheta d^{2}\kappa\label{ChF_red2}
\end{equation}
Inserting now Eqs.~(\ref{ChF_E0}) and (\ref{ChF_tot}) into Eq.~(\ref{ChF_red2}), and
adopting the ${\mathbb{R}}^{4}$ variable $\vec{\varsigma}^{T}\equiv
(\vartheta^{I},-\vartheta^{R},\kappa^{I},-\kappa^{R})$, we obtain
\begin{equation}
\Phi^{(0)}(\vec{\eta}_{a},\vec{\eta}_{b})=\pi^{-2}P_{0}^{-1}\Phi^{tr}%
(\vec{\eta}_{a},\vec{\eta}_{b})\int\exp(-\vec{\varsigma}^{T}\tilde{M}%
\vec{\varsigma}+\vec{v}^{T}\vec{\varsigma})d\vec{\varsigma}\label{ChF_red3}%
\end{equation}
where%
\begin{equation}
\Phi^{tr}(\vec{\eta}_{a},\vec{\eta}_{b})=\exp\left[  -(\vec{\eta}_{a}^{T}%
,\vec{\eta}_{b}^{T})V^{tr}\left(
\begin{array}
[c]{c}%
\vec{\eta}_{a}\\
\vec{\eta}_{b}%
\end{array}
\right)  +i(\vec{d}^{tr})^{T}\left(
\begin{array}
[c]{c}%
\vec{\eta}_{a}\\
\vec{\eta}_{b}%
\end{array}
\right)  \right]  \label{ChF_trace}%
\end{equation}
is the characteristic function of $\rho^{tr}$,%
\begin{equation}
\tilde{M}\equiv\left(
\begin{array}
[c]{cc}%
C+V_{0} & C-\frac{i}{2}\Omega\\
C+\frac{i}{2}\Omega &  C+V_{0}%
\end{array}
\right)
\end{equation}
is a $4\times4$ matrix expressed in terms of the $2\times2$ submatrices
$V_{0}$, $C$ (Charlie's submatrix in Eq.~(\ref{CMtot})) and $\Omega$ (defined in
Eq.~(\ref{Omlabel})),
\begin{equation}
\vec{v}\equiv\left(
\begin{array}
[c]{c}%
2(E^{T}\vec{\eta}_{a}+D^{T}\vec{\eta}_{b})+i(\vec{d}_{0}-\vec{d}_{c})\\
2(E^{T}\vec{\eta}_{a}+D^{T}\vec{\eta}_{b})+i(\vec{d}_{0}-\vec{d}_{c})
\end{array}
\right)
\end{equation}
is an ${\mathbb{R}}^{4}$ vector, with $E$ and $D$ the off-diagonal $2\times2$
submatrices in Eq.~(\ref{CMtot}) and $\vec{d}_{c}$ is the displacement of Charlie's reduced
Gaussian state $\rho_{c}=\mathrm{Tr}_{ab}(\rho)$. Solving the integral in
(\ref{ChF_red3}), we have%
\begin{equation}
\Phi^{(0)}(\vec{\eta}_{a},\vec{\eta}_{b})=P_{0}^{-1}\Phi^{tr}(\vec{\eta}%
_{a},\vec{\eta}_{b})\frac{\exp\left(  \frac{1}{4}\vec{v}^{T}\tilde{M}^{-1}%
\vec{v}\right)  }{\sqrt{g}}\label{ChF_red4}%
\end{equation}
where $g\equiv\det\tilde{M}$ is given in Eq.~(\ref{fattoreG}). Inserting now
Eq.~(\ref{ChF_trace}) in Eq.~(\ref{ChF_red4}), and using $\Phi^{(0)}(\vec{0},\vec
{0})=1$ ($\Longleftrightarrow\mathrm{Tr}_{ab}(\rho^{(0)})=1$), we get
\begin{equation}
\Phi^{(0)}(\vec{\eta}_{a},\vec{\eta}_{b})=\exp\left[  -(\vec{\eta}_{a}%
^{T},\vec{\eta}_{b}^{T})V^{(0)}\left(
\begin{array}
[c]{c}%
\vec{\eta}_{a}\\
\vec{\eta}_{b}%
\end{array}
\right)  +i(\vec{d}^{(0)})^{T}\left(
\begin{array}
[c]{c}%
\vec{\eta}_{a}\\
\vec{\eta}_{b}%
\end{array}
\right)  \right]  \label{ChF_red5}%
\end{equation}
where%
\begin{equation}
\vec{d}^{(0)}=\vec{d}^{tr}+\left(
\begin{array}
[c]{c}%
EM(\vec{d}_{0}-\vec{d}_{c})\\
DM(\vec{d}_{0}-\vec{d}_{c})
\end{array}
\right)  \label{drift_red}%
\end{equation}
is the displacement, with $M$ given in Eq.~(\ref{M}), and the CM $V^{(0)}$ corresponds to the expression of
Eq.~(\ref{CM0}).

\section{Proof of Proposition 3}

Using vectors $\vec{u}^{T}=(u_{x},u_{y})$ in (\ref{u}) and
$\vec{k} (\varphi)^{T}=(k_{x}(\varphi),k_{y}(\varphi))$ in
(\ref{kappa}),\ the fidelity $F(\xi,\varphi)$ can be written as in
(\ref{Fvector}). For an arbitrary $\varphi$, we want to compute
the stationary points $\xi_{\pm}(\varphi)$ of $F(\xi,\varphi)$\ in
the $\xi$ variable. We then introduce the quantity
\begin{equation}
N(\xi,\varphi)\equiv\frac{4\xi^{2}g(\xi,\varphi)^{2}}{F(\xi,\varphi)^{3}}
\frac{d}{d\xi}F(\xi,\varphi)
\end{equation}
so that they are given by $N(\xi,\varphi)=0$ (for
$0<\xi<+\infty$). If the stationary points $\xi_{\pm}(\varphi)$
exist (i.e. they are finite and positive), they have the form
\begin{equation}
\xi_{\pm}(\varphi)=\frac{\omega(\varphi)\pm\sqrt{\Xi(\varphi)}}{\gamma
(\varphi-\pi/2)}\label{CritExpr}
\end{equation}
with
\begin{equation}
\Xi(\varphi)\equiv\omega(\varphi)^{2}+\gamma(\varphi-\pi/2)\gamma(\varphi)
\end{equation}
and $\gamma(\varphi)$ and $\omega(\varphi)$ defined respectively
in (\ref{scalar}) and (\ref{vector}). From the product
\begin{equation}
\xi_{+}(\varphi)\xi_{-}(\varphi)=-\frac{\gamma(\varphi-\pi/2)\gamma(\varphi
)}{\left[  \gamma(\varphi-\pi/2)\right]  ^{2}}\label{product}
\end{equation}
it follows that $\gamma(\varphi-\pi/2)\gamma(\varphi)>0$ implies
the existence of only one stationary point ($\xi_{+}(\varphi)$ and
$\xi_{-}(\varphi)$ have opposite sign). In particular
\begin{equation}
\gamma(\varphi-\pi/2)<0\wedge\gamma(\varphi)<0\Longrightarrow\xi_{-}
(\varphi)>0 ,\label{impl1}
\end{equation}
while
\begin{equation}
\gamma(\varphi-\pi/2)>0\wedge\gamma(\varphi)>0\Longrightarrow\xi_{+}
(\varphi)>0 .\label{impl2}
\end{equation}
Since
\begin{equation}
N(\xi_{\pm}(\varphi)+\varepsilon,\varphi)=\gamma(\varphi-\pi/2)\varepsilon
^{2}\pm2\sqrt{\Xi(\varphi)}\varepsilon
\end{equation}
it is easy to prove that in (\ref{impl1}) $\xi_{-}(\varphi)$ is a
global maximum ($\equiv\xi_{\max}(\varphi)$) while in
(\ref{impl2}) $\xi_{+} (\varphi)$ is a global minimum
($\equiv\xi_{\min}(\varphi)$).

Looking at (\ref{product}), we then analyze the other conditions
$\gamma(\varphi -\pi/2)\gamma(\varphi)<0$ and
$\gamma(\varphi-\pi/2)\gamma(\varphi)=0$.

\begin{itemize}
\item  If $\gamma(\varphi-\pi/2)\gamma(\varphi)<0$ then $\xi_{+}(\varphi)$ and
$\xi_{-}(\varphi)$ are different from zero and have the same sign.
Suppose that $\xi_{\pm}(\varphi)>0$ and consider the case
$\gamma(\varphi -\pi/2)<0\wedge\gamma(\varphi)>0$ (the proof is
analogous in the other case
$\gamma(\varphi-\pi/2)>0\wedge\gamma(\varphi)<0$). From
(\ref{CritExpr}) it follows that $\omega(\varphi)<0$. On the other
hand, from $\gamma
(\varphi)=u_{x}k_{x}(\varphi)+u_{y}k_{y}(\varphi)>0$ and
$\gamma(\varphi
-\pi/2)=u_{x}k_{x}(\varphi-\pi/2)+u_{y}k_{y}(\varphi-\pi/2)<0$, it
follows that
$k_{x}(\varphi)k_{y}(\varphi-\pi/2)-k_{y}(\varphi)k_{x}(\varphi
-\pi/2)=2\omega(\varphi)>0$ which is impossible. For this reason
$\gamma(\varphi-\pi/2)\gamma(\varphi)<0$ implies the non existence
of the stationary points $\xi_{\pm}(\varphi)$. Now, since the
extended function $F(\xi,\varphi)$ surely has a maximum and a
minimum, the global extremal points $\xi_{\min}(\varphi)$ and
$\xi_{\max}(\varphi)$ must be the two border points $0$ or
$+\infty$.
\end{itemize}

The last condition to be analyzed, i.e.
$\gamma(\varphi-\pi/2)\gamma (\varphi)=0$, must be distinguished
in more cases.

\begin{itemize}
\item  First suppose that $\gamma(\varphi-\pi/2)=\gamma(\varphi)=0$. Using
this condition in (\ref{Fvector}), it leads to
$F(\xi,\varphi)=\left[
\det\Gamma^{tr}-k_{x}(\varphi)/k_{y}(\varphi)\right]  ^{-1/2}$,
making the fidelity $\xi-$independent, so that we can freely
choose $\xi_{\min}(\varphi)$ and $\xi_{\max}(\varphi)$ at the
border.

\item  Next suppose that $\gamma(\varphi-\pi/2)\neq0\wedge\gamma(\varphi)=0$.
In such case $\gamma(\varphi)=0\Longrightarrow$
$\omega(\varphi)=-[k_{y} (\varphi)/2u_{x}]\gamma(\varphi-\pi/2)$
and since $k_{y}(\varphi)>0$, $u_{x}>0$ we have that
$\omega(\varphi)$ and $\gamma(\varphi-\pi/2)$ have opposite signs.
On the other hand $\gamma(\varphi)=0\Longrightarrow\xi_{\pm
}(\varphi)=\left[ \omega(\varphi)\pm\left|  \omega(\varphi)\right|
\right] /\gamma(\varphi-\pi/2)$. Consider now
$\gamma(\varphi-\pi/2)>0$ (the proof is analogous in the other
case $\gamma(\varphi-\pi/2)<0$), therefore $\omega(\varphi)<0$ and
we have $\xi_{-}(\varphi)=2\omega(\varphi
)/\gamma(\varphi-\pi/2)=-k_{y}(\varphi)/u_{x}<0$ while
$\xi_{+}(\varphi)=0$, concluding that stationary points do not
exist (global extremal points at the border).

\item  Finally the case $\gamma(\varphi-\pi/2)=0\wedge\gamma(\varphi)\neq0$
can be taken back to the previous ones using periodicity
arguments. Setting $\tilde{\varphi}\equiv\varphi-\pi/2$ in the
last logic proposition, we achieve
$\gamma(\tilde{\varphi})=0\wedge\gamma(\tilde{\varphi}+\pi/2)\neq0$,
from which we can derive both
$\gamma(\tilde{\varphi}-\pi/2)=\gamma (\tilde{\varphi})=0$ and
$\gamma(\tilde{\varphi}-\pi/2)\neq0\wedge
\gamma(\tilde{\varphi})=0$. In either cases we have $\left(
\xi_{\min}
(\tilde{\varphi})=0\wedge\xi_{\max}(\tilde{\varphi})=+\infty\right)
\vee\left(
\xi_{\min}(\tilde{\varphi})=+\infty\wedge\xi_{\max}(\tilde
{\varphi})=0\right)  $. From (\ref{periodicity}) we have
$F(\xi,\tilde
{\varphi})=F(\xi^{-1},\tilde{\varphi}+\pi/2)=F(\xi^{-1},\varphi)$
and therefore $\left(
\xi_{\min}(\varphi)=+\infty\wedge\xi_{\max}(\varphi )=0\right)
\vee\left(  \xi_{\min}(\varphi)=0\wedge\xi_{\max}(\varphi
)=+\infty\right)  $.
\end{itemize}

In conclusion we can summarize all the cases as follows:
\begin{gather}
\gamma(\varphi-\pi/2)\gamma(\varphi)\leq0\Longleftrightarrow\left(
\xi_{\min }(\varphi)=+\infty\wedge\xi_{\max}(\varphi)=0\right)
\vee\left(  \xi_{\min
}(\varphi)=0\wedge\xi_{\max}(\varphi)=+\infty\right) \\
\gamma(\varphi-\pi/2)>0\wedge\gamma(\varphi)>0\Longleftrightarrow\xi_{\min
}(\varphi)=\xi_{+}(\varphi)\wedge(\xi_{\max}(\varphi)=0\vee\xi_{\max}
(\varphi)=+\infty)\\
\gamma(\varphi-\pi/2)<0\wedge\gamma(\varphi)<0\Longleftrightarrow\xi_{\max
}(\varphi)=\xi_{-}(\varphi)\wedge(\xi_{\min}(\varphi)=0\vee\xi_{\min}
(\varphi)=+\infty)
\end{gather}
from which we derive propositions (\ref{sol1}) and (\ref{sol2})
when we consider only the global maximum point
$\xi_{\max}(\varphi)\equiv\bar{\xi
}(\varphi)$.


\begin{thebibliography}{99}
\bibitem{conditional}Note that both fidelities $F^{(n)}$ and $F$ are conditional quantities;
 the first one corresponds to the selection
of a particular outcome $n$ of the measurement, while the second
one does not depend on the outcome but it is determined by the
kind of measurement chosen by Charlie.
\end{thebibliography}
\end{document}